\documentclass[a4paper,10pt]{article}
\usepackage{graphicx}

\begin{document}


\title{On the extragalactic jet asymmetry and composition, and the                					       production of high energy cosmic rays\footnote{Published in NIMA, Vol. 692, p. 280}}

\author{Gizani, Nectaria A.B. $^1$\\
$^1$ Physics Laboratory, School of Science and Technology, Hellenic Open University, \\ Patra, Greece, email: {\tt ngizani@eap.gr} \\}

\date{}
\maketitle

\paragraph{Abstract}

We probe the role of the directional asymmetry between relativistic outflows to kilo-parsec scale jets in the propagation and acceleration of cosmic rays. Our sample contains powerful AGN hosting dense cluster environments. We attempt a thorough description of where and when ultra high energy cosmic ray production and acceleration takes place also using radio observations of the large scale stucture of the AGNs and X-ray data of the hot dense gas in which the AGNs are situated.  

As far as the cosmic ray primaries are concerned, the presence of relativistic protons or mildly relativistic electrons/positrons contributes substantially to the energy budget of the jets enabling them to heat efficiently the intracluster gas. \\


\section{Introduction}

Cosmic rays (CRs) interact with the photons of the Cosmic Microwave Background (CMB). Their proton primary component at energies higher than a few $10^{19}$ eV, loose energy producing photopions. This threshold value sets a theoretical upper limit on the energy of cosmic rays coming from distant sources, known as Greisen-Zatsepin-Kuzmin (GZK) limit (\cite{gzk}). However cosmic rays with energies higher than this cutoff (ultra high energy cosmic rays, or UHECRs) have been observed (e.g. AGASA experiment) and their arrival directions present isotropy. 

The Hillas criterion (\cite{hillas}) taking into account Fermi first order acceleration sets the maximum energy acquired by a charged particle accelerating in a medium containing magnetic field. The energy depends of the magnetic field and its coherent length, the charge of the particle and its velocity. The 'Hillas diagram' suggests a a number of extragalactic sites as origin of UHECRs ($> 10^{20}$ eV) such as active galactic nuclei (eg. \cite{dremmer}; \cite{ensslin}), especially nearby powerful radio galaxies such as Centaurus A (e.g. \cite{go}, \cite{honda}) and M87 (e.g. \cite{hc}, \cite{mb}), radio jets, hotspots, lobes (eg. \cite{dremmer}; \cite{no}), as well as clusters of galaxies. Hardcastle (2010) \nocite{hardcastle} suggests that stochastic particle acceleration of UHECRs to high energies (10$^{20}$~eV) is possible within the large-scale lobes of powerful radiogalaxies as long as the radio sources are at low redshift.  

Particle propagation studies (see for example \cite{olinto}) from the source of the cosmic rays to Earth suggest that if primary UHECR protons travel in straight lines, then within few degrees they should point back to their sources lying in the local Universe and a clear GZK cuttoff should be present. Only large scale intergalactic
magnetic fields affect their propagation significantly unless the Galactic halo magnetic fields are extended. Particles with higher Ze charge however with energies up to 10$^{20}$ eV can be deflected by the Galactic and extragalactic magnetic fields. Faraday rotation measurements, where available, and equipartition arguments imply magnitudes of the intracluster magnetic field of the order of $\mu$G (e.g. \cite{gizanin} and in preparation). Isotropic distribution of heavy nuclei CRs (e.g. iron) favor the Galactic disk as their origin, CR-protons seem to born in extragalactic environments, while photon-primaries point out the early universe.

Spectral energy distribution of AGNs is under reconstruction by invoking leptonic and/or hadronic models for the particle acceleration mechanisms (e.g. \cite{matias}, \cite{orelana}).  

For this work we assume H$_{\circ}$ = 65 km s$^{-1}$ Mpc$^{-1}$ and q$_{\circ}$ = 0 throughout.

\section{Observations of AGN Outflows}

We study 3C\,310 and Hercules A as they present many similarities: They both have double optical nuclei of similar absolute magnitude in R-band at low redshift. They are at the center of cooling flow clusters of galaxies with similar gas temperature. The extended X-ray emission is centered to the  radiogalaxy's core and there is a possible contribution from a point source. They are classified as FR1.5 and have sharply bounded double lobes. Their lobes present asymmetry with respect to brightness, depolarisation and spectral index. They have no compact hotspots. Instead they are probably the only two sources which contain such clear ring-like radio features. Other high-brightness structure is also present with flatter 
spectra than the surrounding diffuse lobes suggesting a renewed outburst from the active nucleus. Their projected B-field follows closely the edges of the rings and lobes. They are both old sources with a steep spectral index (assuming that the flux density is given by S$_\nu \propto \nu^{\alpha}$). Their thermal pressure at the distance of the radio lobes is greater than the lobe minimum pressure. They are both low excitation radiogalaxies.

Their difference are summarised in the following: 3C310 is smaller, and less powerful than Hercules A in the radio, and it also has lower X-ray luminosity.
Her A has weaker radio core. 3C310's core has a flat spectrum, while Her A's steep spectrum. The central electron density of the Hercules A cluster is greater than for the 3C310 cluster. The thermal pressure of the Hercules A cluster is larger meaning that the confinement of its lobes by the intracluster medium is greater

We have observed the parsec scale jets of both sources using VLBI observations at 18 ~cm (\cite{gizanie}, and in preparation). The cores remained unresolved and very weak. However we have discovered that both sources with a well confined and symmetric radio structure, are asymmetric on parsec scales.
If the Her A's outflow observed with the EVN is real, then there is a substantial misalignment of about 35$^{\circ}$ with the collimated and roughly symmetric large scale jets (cf. \cite{gizanir}). Global VLBI radio observations of 3C310 seem to reveal a north-west (NW) emission maybe associated with the pc-scale jet. If this is true then there is a misalignment of 20$^{\circ}$ with respect to the NW kpc-scale jet. Another emission is detected in NE/SW direction misaligned about 100$^{\circ}$ w.r.t. the NW large scale direction. The situation becomes more complicated if we measure the misalignment with respect to the NS kpc-scale jet. The misalignment between the pc-, kpc-jets is a new similarity between Her A and 3C\,310 revealed by our analysis. 

\section{Discussion and Future Work}

Relativistic jets with large scale magnetic fields have large inductive potential and may accelerate protons to ultra high energies. Helical structures are common in AGN jets, arising during jet formation from rotation of magnetized plasma accreting toward the central black hole. Azimuthal electric
currents are likely, yielding a magnetic field component along the jet direction. Our internal to the source magnetic field map (\cite{gizanir}) confirms this. In radiogalaxy jets, boundary layers are often clearly visible in radio polarization. Ostrowski, (2002) has argued that boundary
layers in jets are sites of particle acceleration \nocite{o}.

Hercules A's total radio luminosity is $\sim 4 \times 10^{44}$ coming mostly from the eastern doppler boosted jet. Our VLA observations of the large scale jets (\cite{gizanir}) has shown that  jets are first detectable $\approx$ 2.9 and 3.8 arcsec east and west of the core respectively, and contain knots, helical and rim-brightened features (rings). Jets (eastern--western or jet--counter-jet) are initially collimated with their features roughly symmetric with respect to the core. While the counter jet eventually vanishes after a certain point, the eastern one bends southwards at about 12$^{\circ}$, narrows and eventually fades. Knots and 'rings' are sites of (re)acceleration of relativistic particles. In many places jets flare after a knot as a result of an overpressured strong shocks. The structure of knots suggests shocks moving away from the core through the jet without affecting the underlying material, or they may be moving downstream more slowly than the underlying flow, allowing them to influence the downstream jet. Shocks provide possible inhomogeneity of the internal magnetic field and particles are advected downstream without being able to drift large distances along the shock surface. 

In kpc-scales we have found a dramatic spectral asymmetry between the two sides of the radio emission: The jets and rings have a flatter spectrum than the surrounding lobes and bridge. We interprete this asymmetry between the diffuse lobes as viewing them at different stages of the outburst. Of course, a
detailed spectral ageing analysis should also take into account the
effects of adiabatic expansion on the spectrum. Steep spectral indices found in the lobes of Hercules A and 3C\,310 imply short lifetimes of radiating particles and also re-acceleration of the electrons to some extent. The cooling of the electrons is visible in the steepening of the spectral index of radio emission close to radio galaxies in clusters. These would result in energy redistribution in the ICM.  The energy lost by the energetic particles could be gained by the magnetic fields also heating the ICM. The clusters which are hosted by the two radiosources seem to grow. As a result the energy of CRs in the ICM should increase adiabatically.

Our VLBI observations have suggested a large misalignment between pc- and kpc-scale jets in both powerful, nearby, steep spectrum, ring-like feature containing radio galaxies. For Hercules A this is more obvious as its large scale jets are very collimated (\cite{gizanir}). Not much is known about the curvature of the pc-, kpc-scale jets for steep spectrum sources versus redshift.  The misalignment angle detected is relatively high compared to the ones found in other powerful radio galaxies such as Cygnus A (\cite{Krichbaum}), 3C 109 \cite{Giovannini}. We expect large angles in quasars in the CSS class with steep spectrum cores (e.g. \cite{Fejes} for the OVV quasars 3C 216 and 3C 446 with 'superluminal behaviour', and references therein). This result could be consistent with the suggestion that powerful radio galaxies are the unbeamed counterparts of quasars (\cite{Saripalli}).

Sub-parsec scale acceleration is efficient as long as the scales are comparable to the scale of jet generation or initial collimation.  En$\beta$lin et al., 1996 \nocite{ensslin96} concluded that radio galaxies are powerful enough to heat and support the cluster gas with injected cosmic-ray protons and magnetic field densities (permitted by Faraday rotation and gamma-ray observations of galaxy clusters) within a cluster radius of $\sim$ 1 Mpc. 

ROSAT PSPC and HRI X-ray observations for the Hercules A (\cite{gizanix} ) and 3C\,310 clusters (\cite{hw}) revealed a short cooling time of the emitting cosmic ray electrons. This result combined with the large radio extent of the sources suggest an ongoing acceleration mechanism in the intracluster medium (ICM). Chandra observations of Her A have revealed confinement of the inner jets by the ICM and the presence of X-ray cavities lying within the radiosource \cite{nulsen}. A Chandra proposal for detecting the expected cavities in the 3C\,310 cluster has been submitted. Distortion of the hot X-ray emitting intracluster gas in the form of holes (cavities) is expected, as the radio jets push their way through the gas as they travel in the ICM. The radio/X-ray correlation could either suggest that the relativistic electrons are distributed homogeneously over the lobe, whereas the magnetic field is amplified towards the rim of the lobe region. 

Nulsen et al., 2005 suggest that cavities are not aligned with
the axis of the Hercules A's radio jets and do not contain radio lobes (ghost cavities). 
Detections of several ghost cavities in galaxy clusters emitting in the radio or not - imply a well defined separation of the ICM from the bulk of the relativistic plasma on a timescale of the order of hundred Myr. Leakage can occur however. Benford \& Protheroe, 2008 suggest that if the radio-emitting cosmic-ray clouds are confined within the observable hot gas reservoir of
the cluster,  the dimensions of the X-ray hole (dV) together with gas pressure P, provides a good estimate of the central black hole's energy that is converted
into PdV work against the ICM. Therefore fossil AGN jets and cocoons leave a huge magnetohydrodynamic (MHD) storage even if the AGN is 'dead' \nocite{bp}.

Nulsen at al., 2005 have also detected an enhanced ridge crossing the bright X-ray emmiting region. This emission extends from $\sim 30^{\circ}$ south of east to $\sim 30^{\circ}$ north of west, and forms $\sim 90^{\circ}$ angles to the cavities' axis. The feature forms an angle of $\sim 20^{\circ}$ with the kpc-scale jets and  according to our EVN findings it is towards the direction of the parsec scale jets. This result could suggest a possible correlation between the ridge of X-ray emission and the radio source. The ridge is of nonthermal emission in a relatively cool, dense gas.  The gas may be cool filaments seen in other host sources. 

X-ray cavities contain cosmic rays that later diffuse into larger radio lobes. The latter could be observed either while they are being filled with cosmic rays or shortly after. Mathews \& Brighenti, 2008 \nocite{mb} concluded that initial X-ray cavities will have disappeared by the time that cosmic rays fill the large radio-emitting volume in the cluster gas and a dense filament will be formed.
Therefore combined observations of thermal filaments and radio lobes can be used to trail the propagation of cosmic rays, and study the magnetic structure in the hot ICM and the total cosmic ray energy involved. 

VLA radio maps show that the lobes of both sources are well confined by ICM rather than shocks. Confinement implies that the gas thermal pressure is greater than the minimum pressure in the radio lobes. "Invisible" particles (relativistic protons, low energy  e$^{-}$ / e$^{+}$) should be the dominant particles in the jets and lobes since the internal B is unimportant to the lobe dynamics (\cite{ensslin}; \cite{leahya}). Jets with this particle composition could be energetic sources of cosmic rays.

Boettcher, Dermer \& Finke, 2008 and Dermer et al., 2009, \nocite{bdf} suggested that $\gamma$-ray slowly variable emission in blazars and radiogalaxies could be produced via Compton upscattering of the Cosmic Microwave Background (CMB) photons by shock-accelerated electrons in an extended jet. Boettcher, Dermer \& Finke, 2008  have also shown that such an emission could dominate the bolometric luminosity of the kiloparsec jet if the magnetic fields are of the order of $\sim 10 \mu$G along the jet. Since the value of the magnetic field in both radiosources studied here are of this order, it is within our scope to look for $\gamma$-ray emission from these radio galaxies (FERMI detection)

Following this scientific work we are also going to use analytical hadronic and leptonic models to study particle acceleration. If baryons are indeed present in the jets at sufficient quantities as suspected, then most AGNs exhibiting relativistic jets may be detectable by upcoming neutrino telescopes.


\end{document}